\documentclass[pre,10pt,twocolumn,superscriptaddress,floatfix]{revtex4-2}
\usepackage[utf8]{inputenc}
\usepackage[T1]{fontenc}
\usepackage{hyperref}
\usepackage{amsmath,amsthm,amssymb} 
\usepackage{graphicx}

\usepackage{makecell}
\usepackage{multirow}

\begin{document}
\renewcommand{\acknowledgmentsname}{Acknowledgements}

\title{Spatio-temporal complexity of power-grid frequency fluctuations}

\author{Leonardo~Rydin~Gorj\~ao}
\email{l.rydin.gorjao@fz-juelich.de}
\affiliation{Forschungszentrum J\"ulich, Institute for Energy and Climate Research - Systems Analysis and Technology Evaluation (IEK-STE), 52428 J\"ulich, Germany}
\affiliation{Institute for Theoretical Physics, University of Cologne, 50937 K\"oln, Germany}

\author{Benjamin Sch\"afer}
\email{b.schaefer@qmul.ac.uk}
\affiliation{School of Mathematical Sciences, Queen Mary University of London, United Kingdom}

\author{Dirk~Witthaut}
\email{d.witthaut@fz-juelich.de}
\affiliation{Forschungszentrum J\"ulich, Institute for Energy and Climate Research - Systems Analysis and Technology Evaluation (IEK-STE), 52428 J\"ulich, Germany}
\affiliation{Institute for Theoretical Physics, University of Cologne, 50937 K\"oln, Germany}

\author{Christian Beck}
\email{c.beck@qmul.ac.uk}
\affiliation{School of Mathematical Sciences, Queen Mary University of London, United Kingdom}

\begin{abstract}
Power-grid systems constitute one of the most complex man-made spatially extended structures.  
These operate with strict operational bounds to ensure synchrony across the grid.
This is particularly relevant for power-grid frequency, which operates strictly at $50\,$Hz ($60\,$Hz).
Nevertheless, small fluctuations around the mean frequency are present at very short time scales $<2$ seconds and can exhibit highly complex spatio-temporal behaviour.
Here we apply superstatistical data analysis techniques to measured frequency fluctuations in the Nordic Grid. 
We study the increment statistics and extract the relevant time scales and superstatistical distribution functions from the data.
We show that different synchronous recordings of power-grid frequency have very distinct stochastic fluctuations with different types of superstatistics at different spatial locations, and with transitions from one superstatistics to another when the time lag of the increment statistics is changed.
\end{abstract}

\maketitle

\section{Introduction}

Power-grid systems represent one of the most complex and largest man-made technological structures, which are permanently active and constantly evolving.
These systems supply the power needed for modern society to function.
In order to exchange power between producers and consumers, the grid systems operate as coupled oscillators in strict phase locking, rotating synchronously at a nominal power-grid frequency (e.g. $50\,$Hz or $60\,$Hz).
Understanding the nature of power-grid frequency is crucial as it represents one of the key observables in power system operation and stability as it measures the power balance in the grid: 
It increases in periods of over generation and decreases in periods of scarcity~\cite{Machowski2008}.
If frequency deviation exceeds a threshold dedicated control power plants are ramped up or down to restore the power balance~\cite{Carreras2020}.
At first glance, one could be led to believe that the power-grid frequency is identical across a power grid.
This, however, is not the case, as many studies have indicated~\cite{Schmietendorf2017,Haehne2018,Haehne2019,Wolff2019,Farmer2021}.
Besides oscillations in areas in a power-grid, denoted as intra-area oscillations~\cite{Uhlen2012}, and equivalently across areas in a grid, denoted as inter-area oscillations~\cite{Vanfretti2010}, the presence of other small-scale fluctuations is ubiquitous.
These stochastic fluctuations, i.e., stochastic noise with complex spatio-temporal properties, carry their own physical relevance and dictate particular physical aspects of each location's properties. 
They are influenced by many different factors, such as demand fluctuations, fluctuations in renewable energy production, control actions, trading, and so on.

The spatio-temporal complexity inherent in small frequency fluctuations in power grids is immense, and requires new techniques of analysis to obtain further insight.
In this article we apply superstatistical analysis to this problem, concentrating onto the increment statistics of experimentally measured frequency fluctuations. 
Superstatistical methods, as introduced in Refs.~\cite{Beck2003,Beck2005}, provide a general approach to describe the dynamics of complex non-equilibrium systems with well-separated time scales.
These types of models generate heavy-tailed non-Gaussian distributions by a simple mechanism, namely the superposition of simpler distributions whose relevant parameters are random variables, fluctuating on a much larger time scale. 
Originating in turbulence modelling~\cite{Beck2007}, superstatistics has been applied to many physical systems, such as plasma physics~\cite{Livadiotis2017,Davis2019}, spin systems~\cite{Cheraghalizadeh2021}, cosmic rays~\cite{Yalcin2018,Smolla2020}, self-gravitating systems~\cite{Ourabah2020}, solar wind~\cite{Livadiotis2018}, high-energy scattering processes~\cite{Beck2009,Sevilla2019,Ayala2020}, ultracold gases~\cite{Rouse2017} and non-Gaussian diffusion processes in biophysical systems~\cite{Chechkin2017,Metzler2020,Itto2021}. 
Furthermore, the concept has been successfully applied to other areas as well, for example to environmental time series (e.g. oxygen concentration in rivers)~\cite{Schaefer2021}, wind statistics~\cite{Weber2019}, air pollution~\cite{Williams2020}, bacterial DNA~\cite{Bogachev2017},
financial time series~\cite{Xu2016,Gidea2018}, rainfall statistics~\cite{deMichele2018}, train delays~\cite{Briggs2007}, and plane delays~\cite{Mitsokapas2021}.
In all these cases, an underlying simple distribution, typically Gaussian or exponential, is identified to explain the observed heavy tails of the marginal distributions when integrating over a fluctuating parameter. 
These tails often decay with a power law, although the precise form of the tails depends on the kind of superstatistics considered~\cite{Touchette2005}. 

One of the most commonly used methods to examine the properties of time series data on smallest temporal scales is to study their increment statistics~\cite{Friedrich2011,Tabar2019}, a technique very well-known from turbulent flows~\cite{Castaing1990,Castaing1994,Beck2001a,Beck2007,Beck2005}.
These increments---a set of differences of the time series with given fixed time lag---carry the most fundamental stochastic properties of the underlying noise within the data.
For our particular application, these properties include stochastic fluctuations, their scaling properties, and correlations between fluctuations.
One common aspect is the emergence of heavy tailed statistics in incremental time series, a property well suited to be described by superstatistics.
With respect to power-grid frequency fluctuations, on the one hand we know that two synchronous measurements of the power-grid frequency in the same synchronous region will be basically identical at timescales $>5\!\sim\!10$ seconds.
On the other, recent studies by us~\cite{Schaefer2018,Gorjao2020,Forthcoming} indicate that at scales $<5$ seconds the phase and amplitude synchronisation is not fully achieved, thus recordings of distant locations show local independent properties.

In this paper, we apply the superstatistical approach to unravel the underlying physics of synchronous power-grid frequency recordings, taken at six different sites in the Nordic Grid.
We examine the characteristics of superstatistical properties in incremental time series of the power-grid frequency.
We show that, as observed in Ref.~\cite{Forthcoming}, at scales roughly $>2$ seconds the strong phase locking in coupled power-grid systems dictates the statistical properties of power-grid frequency.
At small timescales, $<2$ seconds, each recording, at a different location, shows distinct superstatistical properties.
We also observe that below $<2$ seconds there is no change of the entropic index $q$, suggesting no change in the superstatistical description in that range.
Generally the southern region of the Nordic Grid exhibits higher entropic indices $q$, indicating that the internal properties of the processes vary greatly.

Our main result, from the superstatistical perspective of the complexity of the power grid, is that the grid consists of different spatial regions with different types of superstatistics, which themselves depend on the time lag chosen for the increment statistics. 
We observe best fits for a combination of lognormal, Gamma (or $\chi^2$), inverse Gamma (or inverse $\chi^2$), and $F$-superstatistics in various spatial regions.
This complexity is higher than in, e.g. isotropic turbulent flows, where usually only one type of superstatistics is observed, which typically is lognormal superstatistics~\cite{Beck2007}.

\begin{figure}[t]
    \includegraphics[width=\linewidth]{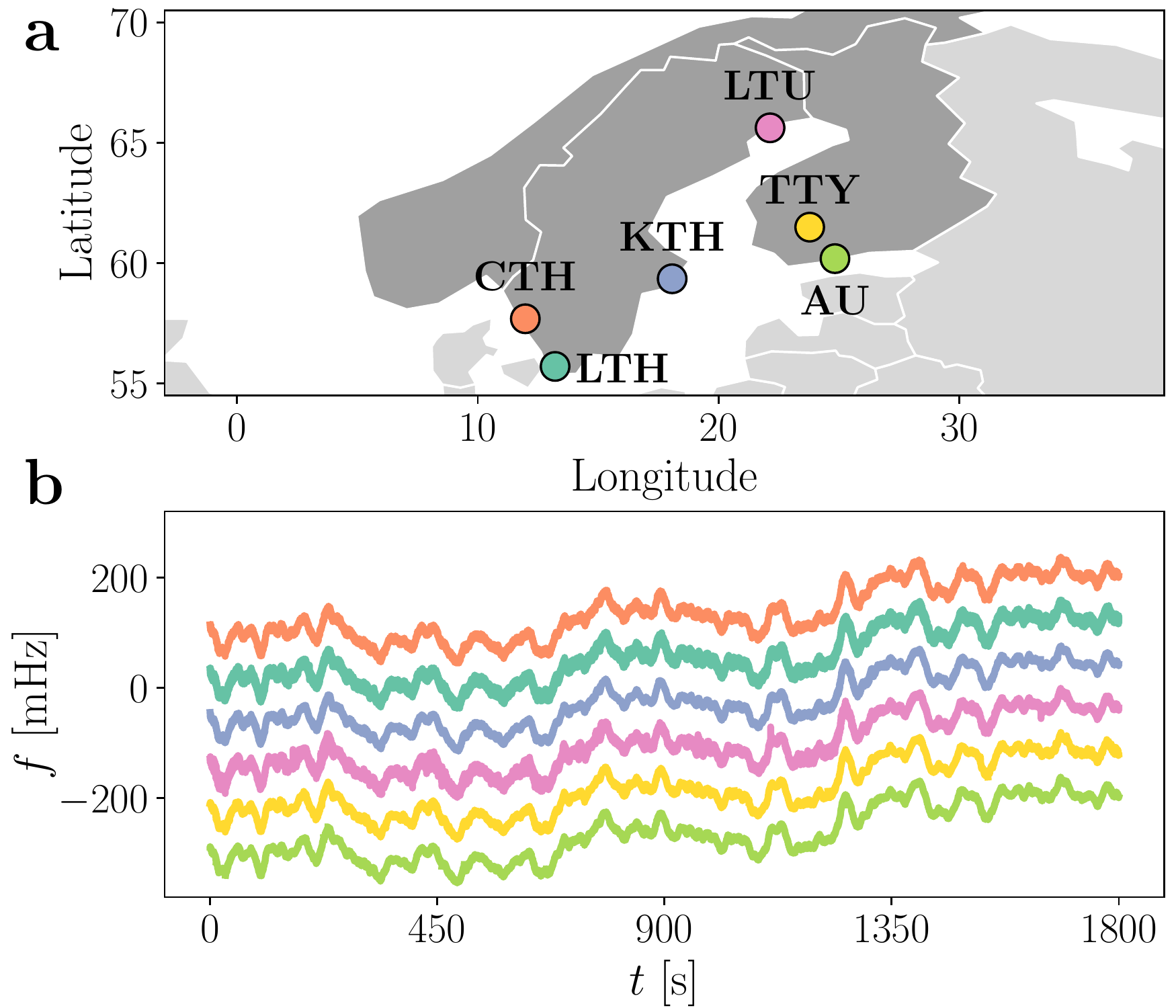}
    \caption{Locations and excerpts of the power-grid frequency recordings. (\textbf{a}) Approximate locations of the recordings across the Nordic Grid: Chalmers University of Technology Gothenburg (CTH); Faculty of Engineering, Lund University (LTH); Royal Institute of Technology Stockholm (KTH); Luleå University of Technology (LTU); Tampere University of Technology (TTY); Aalto University (AU). (\textbf{b}) Excerpts of the recordings in a $30\,$minutes timescale.
    The recordings were taken synchronously and have a time sampling of $0.02\,$s.}\label{fig:trajectories}
\end{figure}

\begin{figure*}[ht]
    \includegraphics[width=\linewidth]{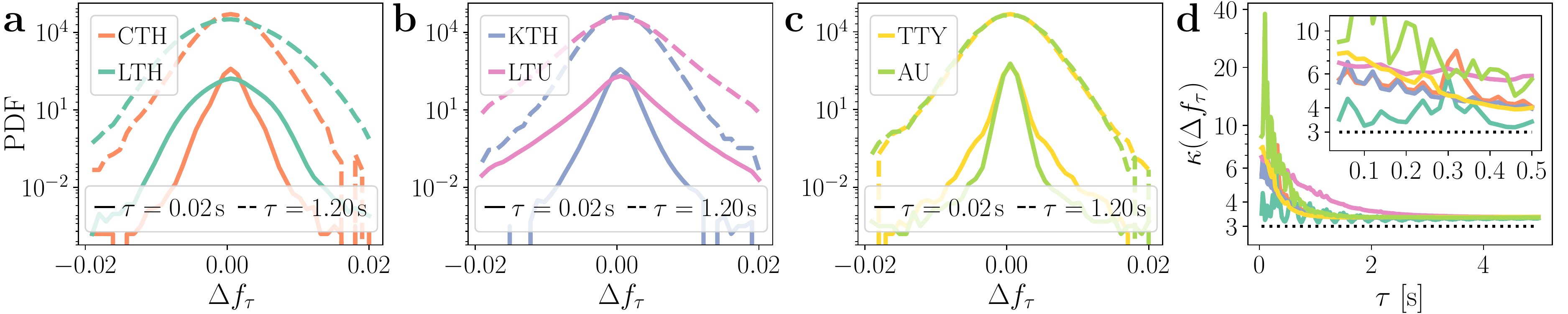}
    \caption{(\textbf{a}-\textbf{c}) Probability density function (PDF) of incremental time series $\Delta f_\tau$ of power-grid frequency recordings at incremental lags $\tau=0.02\,$s and $\tau=1.20\,$s (vertically displaced for clarity).
    (\textbf{d}) Kurtosis $\kappa(\Delta f_\tau)$ of incremental time series with $\tau\in[0.02\,$s$,5\,$s$]$.
    The horizontal dotted line indicates the kurtosis of a normal distribution $\kappa_\mathcal{N}=3$.
    All incremental time series are leptokurtic, i.e. heavy tailed, and their tails become less pronounced as $\tau$ increases.
    Inset shows $\tau\in[0.02\,$s$,0.5\,$s$]$.
    Vertical axis in logarithmic scale. }\label{fig:Increment_dists}
\end{figure*}

\section{Background and methods of analysis}
\subsection{Power-grid frequency dynamics}
In this article we focus on six synchronous power-grid frequency recordings from the Nordic Grid recorded continuously between 21:00 of the 9\textsuperscript{th} to 09:00 of the 11\textsuperscript{th} of September, 2013, with a time sampling of $0.02$ seconds, sufficient to examine the stochastic properties at each location of the recordings in great detail.
The locations of the recordings are indicated by the acronyms of the universities where the recordings were taken: 
Chalmers University of Technology Gothenburg (CTH); Faculty of Engineering, Lund University (LTH); Royal Institute of Technology Stockholm (KTH); 
Luleå University of Technology (LTU); Tampere University of Technology (TTY); Aalto University (AU).
These data come from a former phase-measurement unit network of collaborating universities in the nordic countries, which was in operation between 2012 and 2014~\cite{Almas2014}.

In Fig.~\ref{fig:trajectories}\textbf{a} we show the approximate location of each recording illustrated on a map of the Nordic Grid synchronous area (comprising Norway, Sweden, Finland, and Zealand in Denmark).
In panel \textbf{b} we show excerpts of half an hour of recordings, vertically displaced for clarity.

Synchronous power-grid systems operate at a set frequency to ensure synchrony across the synchronous region---this is the nominal frequency of $50\,$Hz ($60\,$Hz).
From a physical point-of-view, we can understand each location as a node on a complex network of phase-locked oscillators, each with a given inertial mass $M_i$, a damping constant $c_i$, obeying (in a very reductive description)
\begin{equation}\label{eq:swing}
    M_i\frac{\mathrm{d}^2 \theta_i}{\mathrm{d}t^2} = - c_i\frac{\mathrm{d} \theta_i}{\mathrm{d}t} + P_i^\mathrm{m} -\sum_{j\in W}P_j^\mathrm{e},
\end{equation}
with $\theta_i$ the angle of each machine in a co-rotating frame of $W$ interacting machines, $P_i^\mathrm{m}$ the power produced and injected by that machine, and $P_j^\mathrm{e}$ the electrical power extracted by all other machines/loads.
From a more practical point-of-view, we do not have access to this information for all synchronous machines $i$ in a given area.
What we can easily extract from a power-grid is, for example, the frequency $f_i=\mathrm{d} \theta_i/\mathrm{d} t$ (or angular velocity) at a particular location.
This is given, in good approximation, by a Langevin equation~\cite{Hindes2019}
\begin{equation}\label{eq:OU}
    \frac{\mathrm{d} f_i}{\mathrm{d}t} = - \gamma_i f_i + \sigma_i\Gamma_i(t),
\end{equation}
with $\gamma_i$ a friction force (related in some fashion to $c_i$ and $M_i$) and $\Gamma_i(t)$ some noise function, possibly temporally and spatially correlated, with amplitude $\sigma_i$.

Here we are not directly interested in the power-grid frequency, but in the incremental properties of the six measured time series.
The increments of the recordings $\Delta f_\tau(t)$ are given by~\cite{Tabar2019}
\begin{equation}
\Delta f_\tau(t) =  f(t+\tau) - f(t),
\end{equation}
where $\tau$ is the incremental time lag.
In this way we move from a picture in real time $t$ to a ``scale process'' in incremental time lag $\tau$, particularly for small $\tau<5\,$s.
Note here that by examining incremental processes, on these short time scales, one basically excludes the deterministic elements of the process and deals solely with the stochastic characteristics of the fluctuations themselves.
This eliminates concerns about deterministic activity or other long-scale phenomena ($>5\,$s), e.g. dispatch activity, control mechanisms, or power flow changes.

\subsection{Leptokurtic increment statistics}

One of the most important properties to study for
such incremental time series is their probability distribution.
Some previous studies of power-grid frequency involving a stochastic element already remark on the distinct features seen at the level of the increments~\cite{Schmietendorf2017,Haehne2018,Haehne2019,Gorjao2020,Forthcoming}.
Increments statistics often is non-Gaussian, displaying  heavy tails for small time lags, commonly quantified by the kurtosis $\kappa(\Delta f_\tau)$, i.e., the fourth standardised moment, which is given by
\begin{equation}
\kappa(X) = \mathbb{E}\left[\left(\frac{X - \mu_X}{\sigma_X}\right)^4\right] = \frac{\mathbb{E}\left[(X - \mu_X)^4\right]}{\left(\mathbb{E}\left[(X - \mu_X)^2\right]\right)^2},
\end{equation}
where $X$ is a random variable (representing the increments), $\mathbb{E}[\cdot]$ is the expected value, $\mu_X$ is the mean value of $X$, and $\sigma_X^2$ its variance.

In Fig.~\ref{fig:Increment_dists}\textbf{a}-\textbf{c} we display the distributions of the increments at the six locations at the shortest incremental lag $\tau=0.02\,$s and at $\tau=1.20\,$s in a vertical logarithmic scale.
Firstly, we note that rather distinct distributions are immediately evident---some with considerable heavy tails, some less so.
Recall that a normal distribution, in a vertical logarithmic plot, is an inverted parabola.
This is a first, yet clear evidence for the presence of different local stochastic properties in power-grid frequency fluctuations.
There can be numerous reasons for this: characteristics of the local generation, e.g. renewable energy generation, nuclear, or fossil fuel, differences of local consumption 
patterns, distant consumption requiring transferring power over transmission lines, and so on.

In order to adequately quantify the heavy-tailedness of these incremental time series we examine the kurtosis $\kappa$ as a function of the incremental lag $\tau$, in a similar way as
this is done in turbulent flows~\cite{Beck2001b}.
In Fig.~\ref{fig:Increment_dists}\textbf{d} we show the kurtosis $\kappa(\Delta f_\tau)$ of the incremental time series $\Delta f_\tau$ with $\tau\in[0.02\,$s$,5\,$s$]$.
The horizontal dotted line indicates the kurtosis of a normal distribution, i.e., $\kappa_\mathcal{N}=3$.
One can clearly observe all recordings exhibit $\kappa(\Delta f_\tau)>3$ for all $\tau$, and as $\tau$ increases, the kurtosis of the increments tends to $\kappa(\Delta f_{\tau\gg0}) \approx3.35$.
Distributions with large kurtosis, i.e., $\kappa>3$, are called leptokurtic (conversely platykurtic).
Similar phenomena were seen for various other power-grid frequency recordings~\cite{Schaefer2018,Gorjao2020}, yet a clear explanation why the incremental time series does not fully relax to a normal distributions with $\kappa = 3$ on a large scale is currently absent.

In the following section we will employ superstatistics as the possible mechanism to explain the presence of leptokurtic incremental time series.
Such a superstatistical approach offers an explanation for the observed leptokurtic probability distributions.
However, for this to work one has to carefully examine if time scale separation is realised for power-grid frequency fluctuations, such that a superstatistical treatment is justified.

\subsection{Superstatistical generation of leptokurtic distributions}

We start from the usual assumption of superstatistics that the underlying equilibrium state of the increment statistics is a simple distribution, i.e. Gaussian, on a suitable time scale $T$ to be determined.
This agrees with our formulation of the power-grid frequency dynamics as a Langevin equation in~\eqref{eq:OU}.

The increment statistics of the entire
time series can then be viewed as a superposition of Gaussian distributions with different variances, weighted  via a scaling function $f(\beta)$, in itself a normalised probability distribution, such that the probability density function $p(\Delta f_\tau)$ of the increments $\Delta f_\tau$ is given by
\begin{equation}\label{eq:fund_super_dist}
    p(\Delta f_\tau) = \int_{0}^{\infty} f(\beta) p_\mathcal{N}(\Delta f_\tau|\beta) \mathrm{d}\beta,
\end{equation}
with 
\begin{equation}
    p_\mathcal{N}(\Delta f_\tau|\beta)  =  \sqrt{\frac{\beta}{2\pi }}e^{-\frac{1}{2}\beta \Delta f_\tau^2}
\end{equation}
i.e., a normal distribution dependent on a scaling parameter $\beta$, regarded as a volatility.

What underlies the description of a superposition of Gaussian distribution has a clear physical explanation: The process' internal characteristics change in time.
This is not hard to imagine. The amount of power generation and consumption changes over the day, and so does the contribution of each type of energy source, the total inertia of the system (linked to the number of conventional generator connected to the power grid), amongst many other properties of generation, consumption, and power transport.

Two points are crucial here: firstly, the superstatistical distribution $f(\beta)$ can be different at each spatial location, as we will see.
Secondly, in principle, the shape of $f(\beta)$ can change for different incremental lags $\tau$.
The probability density function $f(\beta)$ can generally be any appropriate and normalised distribution with support in $[0,\infty)$.
We will subsequently discuss four candidate distributions for $f(\beta)$: the log-normal distribution, the Gamma distribution (or $\chi^2$ distribution), the inverse-Gamma distribution (or inverse $\chi^2$ distribution), and the $F$ distribution, in line with what was done in Ref.~\cite{Beck2003}.
In our analysis, we will also show that it is hard to
extract a particular distribution $f(\beta)$ in a unambiguous way, i.e., many distributions are consistent with the data. 

The procedure to extract $f(\beta)$ from a given
data set is as follows. Under the assumption that different scales exist in the incremental time series of power-grid frequency recordings, one segments the data into small
sub-slices, called `snippets' in the following.
Then one studies each snippet's probability distribution.
Particularly, one examines each snippet of the incremental time series---for $\tau=0.04\,$s to begin with---and extracts its kurtosis, which we then average for all snippets with a window of length $\delta t$
\begin{equation}
    \kappa_{\delta t}(\Delta f_\tau) = \left\langle \frac{\frac{1}{\delta t}\sum_{i=(j-1)\delta t +1}^{j\delta t} {\Delta f_\tau}_i^4}{(\frac{1}{\delta t}\sum_{i=(j-1)\delta t +1}^{j\delta t} {\Delta f_\tau}_i^2)^2} \right\rangle_{\delta t}
\end{equation}

with $\langle \cdot \rangle_{\delta t}$ the average over each snippets' length.

\begin{figure}[t]
    \includegraphics[width=\linewidth]{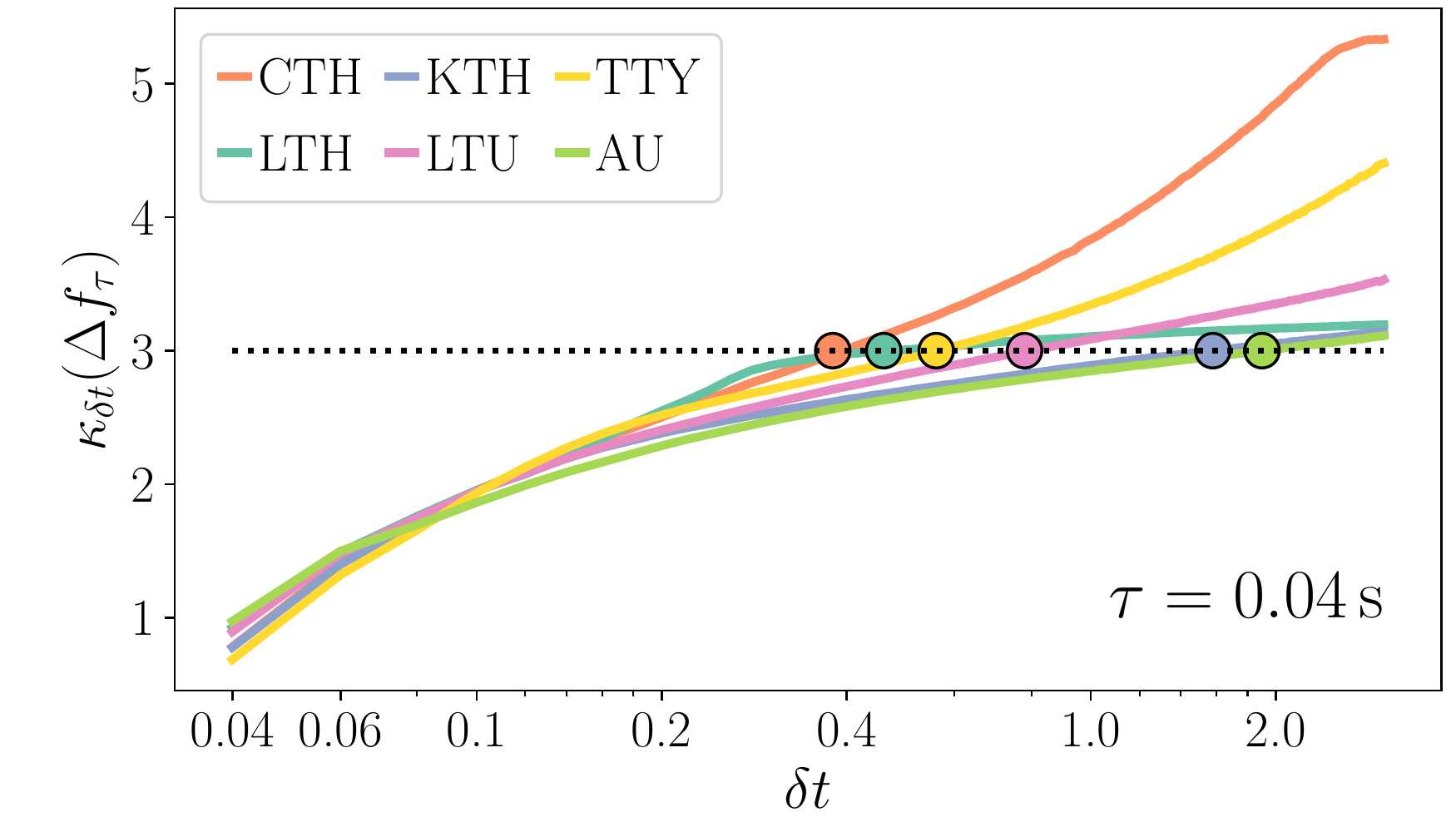}
    \caption{Obtaining the long superstatistical time $T$ for the incremental time series $\Delta f_\tau$ with $\tau=0.04\,$s.
    Snippets of size $\delta t$ of each incremental time series are taken and the average kurtosis is obtained.
    The long superstatistical time $T$ is obtained at the crossing where $\kappa=3$, i.e., the distributions are normal (horizontal dotted line).
    Each circle indicates this crossing, with respective long superstatiscal times $T$ for each location being
    $T_{\mathrm{CTH}} = 0.38\,$s, $T_{\mathrm{LTH}} = 0.46\,$s, $T_{\mathrm{KTH}} = 1.58\,$s, $T_{\mathrm{LTU}} = 0.78\,$s, $T_{\mathrm{TTY}} = 0.56\,$s, $T_{\mathrm{AU}} = 1.90\,$s}\label{fig:crossing} 
\end{figure}

For very short snippets the average snippet kurtosis $k(\delta t)$ is smaller than $3$, i.e., the average distribution shows a lack of tailedness (it is platykurtic).
As the snippets increase in size $\delta t$, the average kurtosis grows.
As we have seen in Fig.~\ref{fig:Increment_dists}\textbf{d}, in the limit of the snippet being equal to the entire time series, the kurtosis is larger than $3$, thus there is a certain snippet size---at the long superstatistical time $T$---at which the average snippet kurtosis is $3$
\begin{equation}
    \kappa_{\delta t}(\Delta f_\tau) \equiv 3,
\end{equation}
and thus the snippets are, on average, Gaussian distributed
for this particular time scale.

In Fig.~\ref{fig:crossing} the procedure of finding the long superstatistical time $T$ is illustrated, for the incremental time series with $\tau=0.04\,$s.
The circles indicate the crossing at which the snippets are normally distributed, from which we determine the long superstatistical time $T$.
Notice here that this long superstatistical time $T$ is depending on the location where the time series is measured.
It also varies for different incremental lags $\tau$.

Having obtained the superstatistical timescale $T$ for each time series (see Fig.~\ref{fig:crossing}), i.e., the snippet length at which the average snippet kurtosis is $3$, one can subsequently extract the distribution of the scaling function $f(\beta)$ by extracting the inverse of the variance of each snippet at $\delta t = T$
\begin{equation}
    \beta_T(t) = \frac{1}{\langle \Delta f_\tau^2 \rangle_{T} - \langle \Delta f_\tau \rangle_{T}^2}.
\end{equation}
We thus get a distribution of values for $\beta_T(t)$ from which we determine the scaling function $f(\beta)$ by simply finding its distribution, i.e., examining its histogram.
Notice that if all snippets have the same variance then the function $f(\beta)$ is very narrow (in the limit it is a single point, i.e., a Dirac delta function).
This means that there are no changes of $\beta$ and the internal characteristics of the time series remains the same over time.
If the variance of each snippet varies, one obtains a distribution, which is given by $f(\beta)$.

\begin{figure}[t]
    \includegraphics[width=\linewidth]{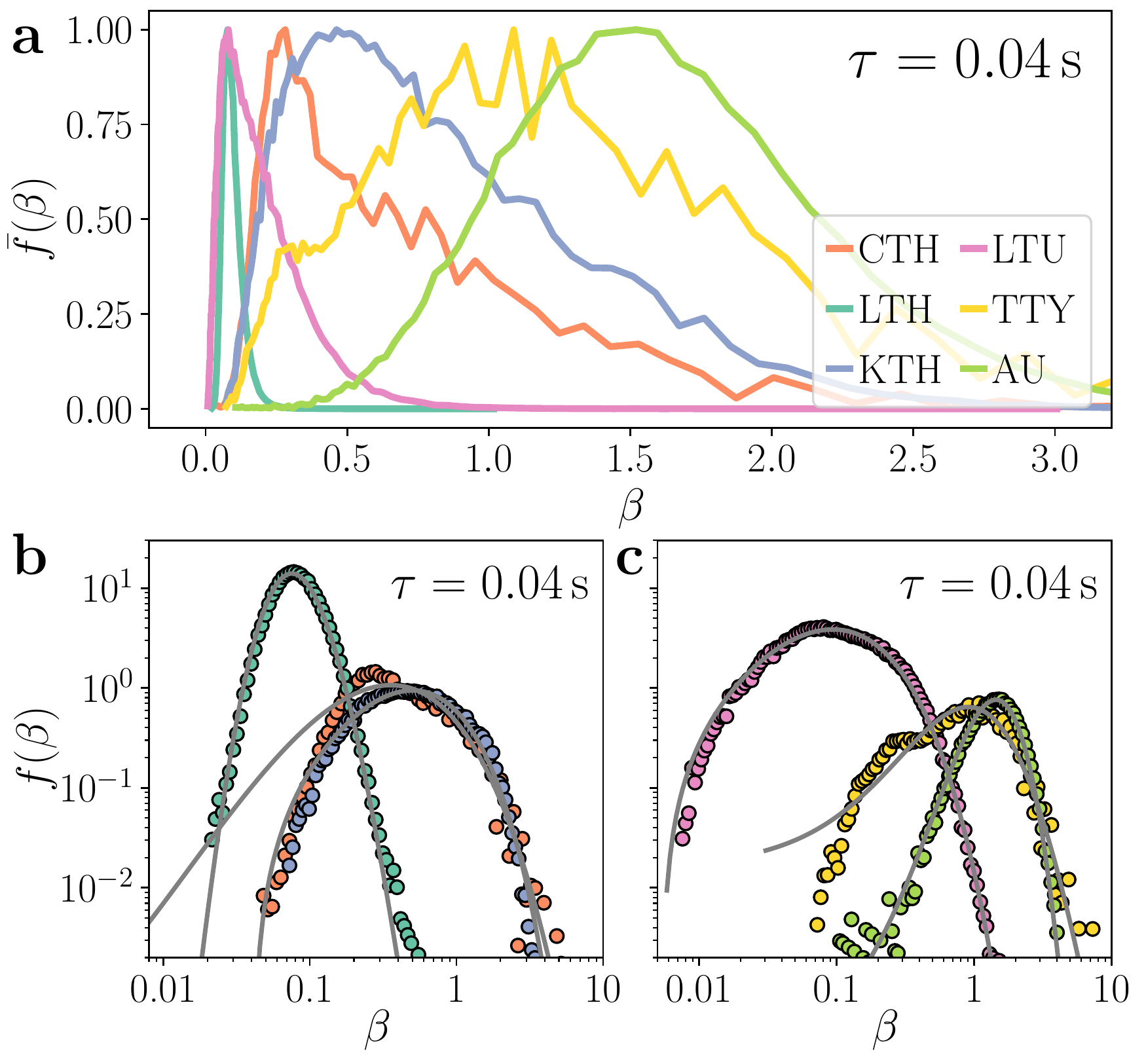}
    \caption{Scaling function $f(\beta)$ of incremental time series at incremental lag $\tau=0.04$\,s.
    (\textbf{a}) displays a shrunk scaling function with a maximum value of $1$ for visual comparison ($\bar{f}(\beta)=f(\beta)/\operatorname{max}(f(\beta))$).
    (\textbf{b}-\textbf{c}) display the scaling function $f(\beta)$ in a double logarithmic scale.
    The best fitting distribution is displayed for each location according to the minimisation of the Kolmogorov--Smirnov distance $D_n$ in \eqref{eq:ks_D}, as found in Tab.~\ref{table}.
    In CTH: Gamma; in LTH: $F$; in KTH and LTU: Gamma; in TTY and AU: inverse Gamma.
    Distributions obtained for $\tau=0.04\,$s.
    The superstatiscal times $T$ for each location are
    $T_{\mathrm{CTH}} = 0.38\,$s, $T_{\mathrm{LTH}} = 0.46\,$s, $T_{\mathrm{KTH}} = 1.58\,$s, $T_{\mathrm{LTU}} = 0.78\,$s, $T_{\mathrm{TTY}} = 0.56\,$s, $T_{\mathrm{AU}} = 1.90\,$s}\label{fig:beta}
\end{figure}

In Fig.~\ref{fig:beta} the underlying distributions $f(\beta)$ are extracted from the incremental time series at incremental lag $\tau=0.04\,$s~\cite{Beck2005}, simply by doing a histogram of $\beta$ as observed in each time slice.
In panel \textbf{a} we display the distributions of $\bar{f}(\beta)=f(\beta)/\operatorname{max}(f(\beta))$, which have been rescaled such that each distributions' peak has a maximum value $1$, so that they are visually comparable.
We immediately see different widths for $f(\beta)$ for the different locations.

As stated, we do not assume \textit{a priori} a specific distribution $f(\beta)$---on the contrary, we wish to find this just from the distribution of the volatilities $\beta$
in our given data set.
Note that many theoretical distributions can be compatible with the data.
We will show that many suggested distributions are compatible,
although the Kolmogorov Smirnov test
will point to a particular one as having
least distance. One then sees that different location have different $f(\beta)$ distributions.

\begin{table}[t]
  \centering
  \begin{tabular}{r|c|c|c|c|c}
    \hline
    Location & $T$ & $f_{\mathrm{log}\mathcal{N}}(\beta)$ & $f_{\Gamma}(\beta)$ & $f_{\mathrm{inv}\Gamma}(\beta)$ & $f_{F}(\beta)$ \\
    \hline
    \multirow{2}{*}{CTH} & \multirow{2}{*}{0.38\,s} & 0.0344 & \textbf{0.0326} & 0.0473 & 0.0384 \\
    & & 5.5\% & \textbf{0.0}\% & 45.0\% & 17.7\% \\ \hline
    \multirow{2}{*}{LTH} & \multirow{2}{*}{0.46\,s}  & 0.0178 & 0.0988 & 0.0081 & \textbf{0.0080} \\
    & & 124.4\% & 1141.9\% & 1.3\% & 0.0\% \\ \hline
    \multirow{2}{*}{KTH} & \multirow{2}{*}{1.58\,s} & 0.0287 & \textbf{0.0180} & 0.0319 & 0.0184 \\
    & & 59.4\% & \textbf{0.0}\% & 77.2\% & 2.1\% \\ \hline
    \multirow{2}{*}{LTU} & \multirow{2}{*}{0.78\,s} & 0.0251 & \textbf{0.0121} & 0.0299 & 0.0147 \\
    & & 106.8\% & \textbf{0.0}\% & 146.7\% & 1.2\% \\ \hline
    \multirow{2}{*}{TTY} & \multirow{2}{*}{0.56\,s} & 0.0249 & 0.0342 & \textbf{0.0239} & 0.0240 \\
    & & 4.2\% & 43.1\% & \textbf{0.0}\% & 0.4\% \\ \hline
    \multirow{2}{*}{AU} & \multirow{2}{*}{1.90\,s} & 0.0023 & 0.0048 & \textbf{0.0016} & 0.0024 \\
    & & 43.5\% &  198.0\% & \textbf{0.0}\% & 46.7\% \\
    \hline
  \end{tabular}
\caption{Kolmogorov--Smirnov distance $D_n$ (upper value), given by~\eqref{eq:ks_D}, and relative percentual difference from smallest $D_n$ (lower value) for volatilities $\beta$ at incremental lag $\tau=0.04$\,s.
The long superstatistical time $T$ for each location is also indicated.
Four candidate distributions are fitted with a maximum likelihood estimation: log-normal $f_{\mathrm{log}\mathcal{N}}(\beta)$~\eqref{eq:dist_lognormal}, Gamma $f_{\Gamma}(\beta)$~\eqref{eq:dist_gamma}, inverse Gamma $f_{\mathrm{inv}\Gamma}(\beta)$~\eqref{eq:dist_inverse_gamma}, and F $f_{F}(\beta)$~\eqref{eq:dist_F}.
In bold are indicated the smallest Kolmogorov--Smirnov distances $D_n$ for each location.
The percentual values are calculated via $\left(D_n - \mathrm{min}(D_n)\right)/\mathrm{min}(D_n)$ for each location.}\label{table}
\end{table}

An interesting result is the fact that the shape of the
distribution $f(\beta)$ is influenced by the time lag
of the increment statistics. Accordingly, at larger time lags one obtains different optimal fits. This is
shown in Tab.~\ref{table_2}.

\begin{table}[h]
  \centering
  \begin{tabular}{r|c|c|c|c|c}
    \hline
    Location & $T$ & $f_{\mathrm{log}\mathcal{N}}(\beta)$ & $f_{\Gamma}(\beta)$ & $f_{\mathrm{inv}\Gamma}(\beta)$ & $f_{F}(\beta)$ \\
    \hline
    \multirow{2}{*}{CTH} & \multirow{2}{*}{0.24\,s}     & \textbf{0.0432}   & 0.0968 & 0.0545 & 0.0521 \\
                                                        & & \textbf{0.0}\%    & 123.9\% & 26.0\% & 20.4\%  \\ \hline
    \multirow{2}{*}{LTH*} & \multirow{2}{*}{0.26\,s}     & 0.0372   & 0.0591 & 0.0202 & \textbf{0.0200}   \\
                                                        & & 85.8\%   & 195.3\% & 0.9\% & \textbf{0.0}\%  \\ \hline
    \multirow{2}{*}{KTH*} & \multirow{2}{*}{112.5\,s}     & \textbf{0.0222}   & 0.0314 & 0.0230  & 0.0228 \\
                                                        & & \textbf{0.0}\%    & 41.3\% & 3.6\% & 2.6\%  \\ \hline
    \multirow{2}{*}{LTU} & \multirow{2}{*}{0.6\,s}     & 0.0454   & 0.0629 & 0.049  & \textbf{0.0317} \\
                                                        & & 43.1\%   & 98.5\% & 54.6\% & \textbf{0.0}\%  \\ \hline
    \multirow{2}{*}{TTY} & \multirow{2}{*}{0.5\,s} & 0.0216   & 0.0243 & 0.0235 & \textbf{0.0168} \\
                                                        & & 28.5\%   & 44.5\% & 40.1\% & \textbf{0.0}\%  \\ \hline
    \multirow{2}{*}{AU} & \multirow{2}{*}{20.84\,s}   & 0.0103   & \textbf{0.0047} & 0.0137 & 0.0052 \\
                                                        & & 118.3\%  & \textbf{0.0}\% & 189.8\% & 10.8 \% \\ 
    \hline
  \end{tabular}
\caption{Kolmogorov--Smirnov distances $D_n$ (upper value), given by~\eqref{eq:ks_D}, and relative percentual difference from smallest $D_n$ (lower value) for volatilities $\beta$ at incremental lag $\tau=1.20$\,s.
The two locations LTH and KTH are not examined in Sec.~\ref{sec:III} due to their coupling, leading to large variations in each recordings' kurtosis.
Note the differences of optimal fits as compared to Tab.~\ref{table},
which is for a much smaller time lag.}\label{table_2}
\end{table}

We consider here four candidate distributions.
A log-normal distribution $f_{\mathrm{log}\mathcal{N}}(\beta)$ of two parameters $s>0$ and $\mu$ with probability density function given by
\begin{equation}\label{eq:dist_lognormal}
    f_{\mathrm{log}\mathcal{N}}(\beta) = \frac{1}{\sqrt{2\pi}s\beta}\mathrm{exp}{\left(-\frac{\left(\ln{\beta}-\mu\right)^2}{2s^2}\right)},
\end{equation}
with $\beta\in(0,\infty)$.
A Gamma distribution $f_{\Gamma}(\beta)$ of two parameters $b>0$ and $c>0$ with probability density function given by
\begin{equation}\label{eq:dist_gamma}
    f_{\Gamma}(\beta) = \frac{1}{b\Gamma(c)}\left(\frac{\beta}{b}\right)^{c-1}\!\!\!\!\!\!\!\!\mathrm{exp}{\left(-\frac{\beta}{b}\right)},
\end{equation}
with $\beta\in(0,\infty)$ and $\Gamma$ the Gamma function.
An inverse Gamma distribution $f_{\mathrm{inv}\Gamma}(\beta)$ of two parameters $b>0$ and $c>0$ with probability density function given by
\begin{equation}\label{eq:dist_inverse_gamma}
    f_{\mathrm{inv}\Gamma}(\beta) = \frac{b^c}{\Gamma(c)}\frac{1}{\beta^{c+1}}\mathrm{exp}{\left(-\frac{b}{\beta}\right)},
\end{equation}
with $\beta\in[0,\infty)$.
Lastly, we consider as well an $F$ distribution $f_{F}(\beta)$ of three parameters, $v$ and $w$ positive integers and $b>0$, with probability density function given by
\begin{equation}\label{eq:dist_F}
    f_{F}(\beta) = \frac{\Gamma ((v+w)/2)}{\Gamma (v/2) \Gamma (w/2)} \left( \frac{bv}{w} \right)^{v/2}\!\!\!\!\!\!\!\! \frac{\beta^{v/2-1}}{\left(1+ \frac{vb}{w}\beta\right)^{(v+w)/2}},
\end{equation}
with $\beta\in[0,\infty)$.

Naturally the subsequent question is how to assess which distribution best fits the volatilities $\beta$ we extracted.
The Kolmogorov--Smirnov test is a non-parametric test that evaluates the equality of two continuous distribution functions via their cumulative density functions (CDF).
The Kolmogorov--Smirnov distance gives the maximal difference between the empirical cumulative distribution function $F_n(\beta)$ (for us given by the the CDF of the empirical $f(\beta)$) and a chosen cumulative distribution function $F(\beta)$ (one of our four candidate distributions).
It is given by
\begin{equation}\label{eq:ks_D}
    D_n= \sup_\beta |F_n(\beta)-F(\beta)|,
\end{equation}
with $\sup$ denoting the supremum.
We do not have access to the true cumulative distribution function $F_n(\beta)$ thus we employ a numerical maximum likelihood estimation for each of the four aforementioned distributions~\eqref{eq:dist_lognormal}, \eqref{eq:dist_gamma}, 
\eqref{eq:dist_inverse_gamma}, and~\eqref{eq:dist_F}.

In Tab.~\ref{table} we display the Kolmogorov--Smirnov statistics $D_n$ in \eqref{eq:ks_D} for each location and all four distributions for the incremental lag $\tau=0.04$\,s.
We note two things: Different distributions are a best fit for different locations, but often the Kolmogorov--Smirnov distance $D_n$ is considerably small for another distribution as well.
The highlighted numbers indicate the smallest Kolmogorov--Smirnov distances $D_n$.
These are compared percentage-wise by taking the difference of the Kolmogorov--Smirnov distances $D_n$ divided by the smallest $D_n$ of each location.
Whereas in CTH a Gamma distribution minimises the Kolmogorov--Smirnov distance $D_n$, in LTH the $F$-distribution is better suited, in KTH and LTU a Gamma distribution is again suited best and lastly in TTY and AU the inverse Gamma distribution comes first.
Apart from AU, to which we will return later, there is always another distribution which would fit the empirical distribution $f(\beta)$ with a comparably small Kolmogorov--Smirnov distance $D_n$.
In other words, the best theoretical model cannot be identified unambiguously.

We also note that the best-fitting distribution can change if the incremental time lag $\tau$ is changed, e.g. compare Tab.~\ref{table} with Tab.~\ref{table_2}.
This is not surprising for two reasons: firstly, the strong coupling between the locations can influence the statistics of each location. 
Secondly, the data itself is limited to $36$ hours of continuous measurements, which is not sufficient to uncover with exactness the underlying statistics, if there truly is a single one.

We thus conclude that that there is some ambiguity to identify the precise particular form of superstatistics from the given data.
This nevertheless does not prevent us from examining in detail intrinsic properties of the incremental time series as a function of the incremental lag $\tau$, as we will do in the next section.

\section{Superstatistical properties as a function of the incremental time lag}\label{sec:III}

In order to ensure that a superstatistical description is viable, we need to determine whether a given time series achieves a local equilibrium at a scale much smaller than the superstatistical variation time scale of the parameter $\beta$, given by the long superstatistical time $T$.
To do so, we examine the relaxation time of the correlation function of the incremental time series.
Recall that the auto-correlation function is given by
\begin{equation}
    C(t-t') = \mathbb{E}[(X(t)-\mu_X)(X(t')-\mu_X)],
\end{equation}
where for our case here $X(t)=\Delta f_\tau(t)$ (or $X(t)=\beta(t)$), i.e., we examine the auto-correlation function of the incremental time series $\Delta f_\tau(t)$ (or of the volatilities $\beta(t)$).
Assuming that the correlation function initially decays exponentially for the incremental time series, we can extract the decay time $\rho$ of the exponential relaxation, i.e., $\rho$ such that $C(d) = e^{-1}C(0)$, which dictates the short superstatistical time $d$.

\begin{figure}[t]
    \includegraphics[width=\linewidth]{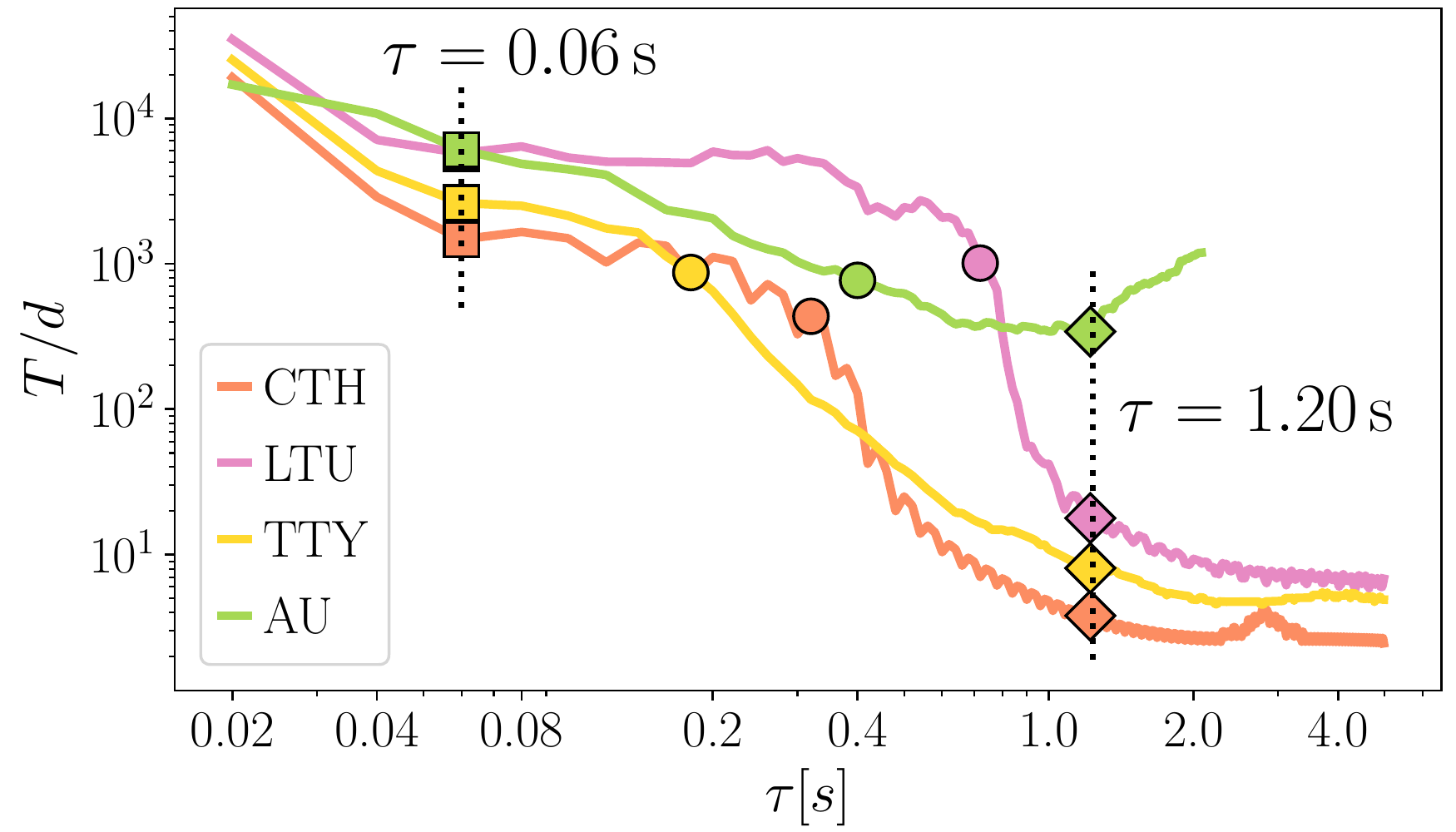}
    \caption{Ratio of superstatistical times $T/d$ (long over short superstatistical times) for increasing incremental lags $\tau$, in a double logarithmic scale.
    For short incremental lags $\tau<1.2\,$s the ratio $T/d$ is large, ensuring a superstatistical description is adequate.
    Noticeably is the fact that $T/d$ decreases for almost all incremental time series, with the exception of AU.
    The vertical lines, at $\tau=0.06\,$s and $\tau=1.20\,$s, combined with the four circles (\raisebox{-0.5mm}{\scalebox{1.5}{$\circ$}}) indicated the incremental lag at for which we display scale functions $f(\beta)$ found in Fig.~\ref{fig:more_betas}.
    Circles (\raisebox{-0.5mm}{\scalebox{1.5}{$\circ$}}) are $\tau_{\mathrm{CTH}} = 0.30\,$s, $\tau_{\mathrm{LTU}} = 0.70\,$s, $\tau_{\mathrm{TTY}} = 0.16\,$s, and $\tau_{\mathrm{AU}} = 0.38\,$s.
    LTH and KTH are excluded due to  their strong coupling resulting in abnormal variations of the kurtosis of the incremental lags.}\label{fig:Ts}
\end{figure}

To ensure a superstatistical description is possible, the short superstatistical time needs to be smaller than the large superstatistical time $d\ll T$, as the names suggest~\cite{vanderStraeten2009}. 
This guarantees that, locally, each incremental time series reaches an equilibrium before the larger scale superstatistics changes the physics of the process.
In Fig.~\ref{fig:Ts} we examine this relation for varying incremental lags $\tau$, confirming that for incremental lags $\tau\lesssim1.2\,$s the relation $d\ll T$ holds.
We mark here a transition time at roughly $\tau\approx1.2\,$s as the starting incremental lag where superstatistics loses validity, with the clear exception of the recordings at AU.
We note that in another work by us~\cite{Forthcoming} we show that this is roughly the same scale where the power-grid frequency increments at different location lose their independence and their phases become effectively identical to each other.
This seems to be simultaneously the scale where the superstatistical modelling loses validity.
Interestingly, the ratio of statistical times $T/d$ decreases with $\tau$ (with the exception of AU), which appears to be in contrast to what is observed in turbulent flows~\cite{Beck2005}.

\begin{figure*}[t]
    \includegraphics[width=\linewidth]{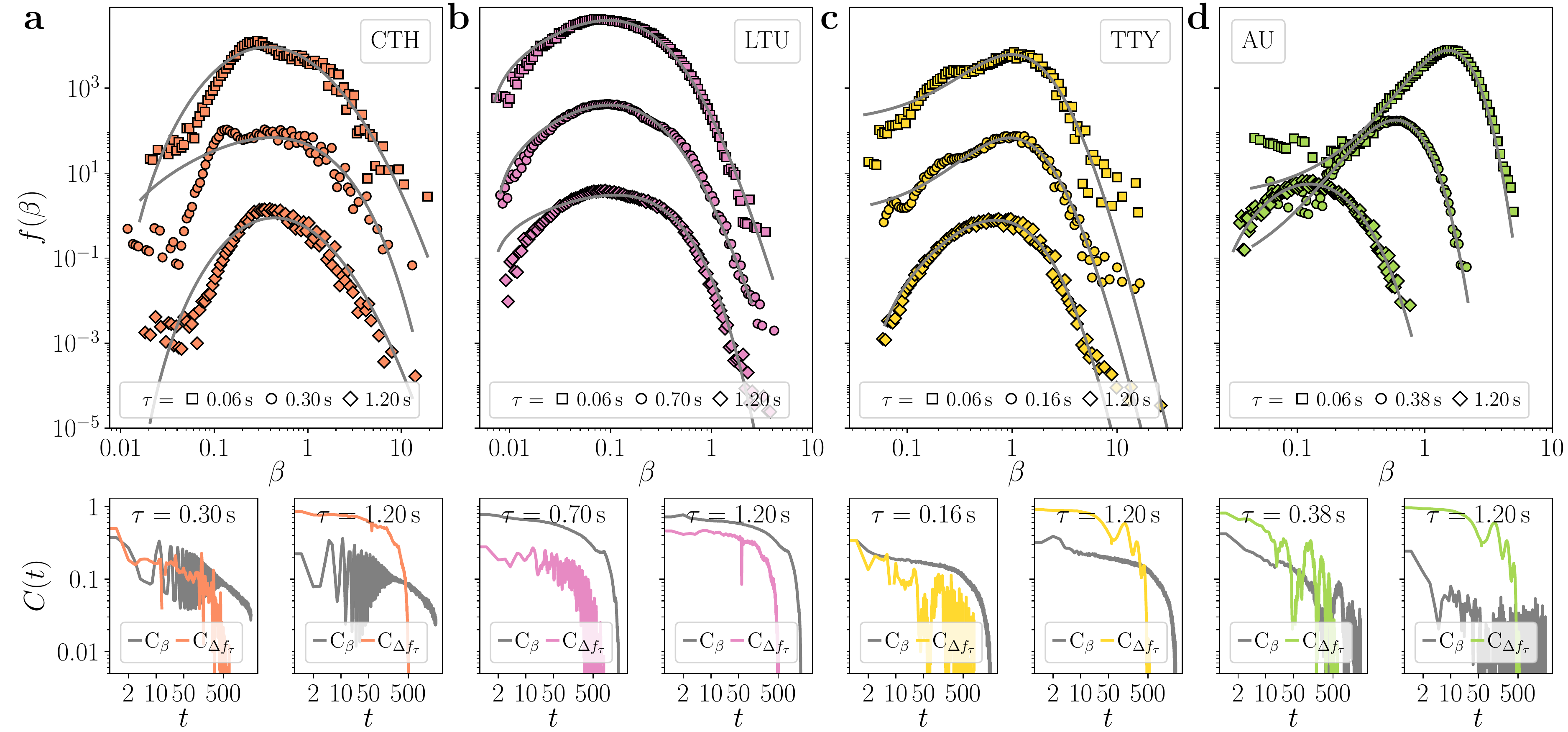}
    \caption{Underlying scaling function $f(\beta)$ of incremental time series for
    \textbf{a} CTH, \textbf{b} LTU, \textbf{c} TTY, and \textbf{d} AU, for incremental lags $\tau=0.06\,$s ($\Box$ top) and $\tau=1.20\,$s (\rotatebox[origin=c]{45}{$\Box$} bottom).
    The delay choices of $\tau$ are indicated in Fig.~\ref{fig:Ts} by the two vertical dashed lines and the four circles (\raisebox{-0.5mm}{\scalebox{1.5}{$\circ$}}).
    The distributions are vertically displaced for clarity.
    Solid lines indicate the best fitting distributions.
    The bottom panel displays for each location the correlation function of the incremental time series $C_{\Delta f_\tau}$ and the correlation function of the volatility $\beta(t)$, $C_\beta$, for the two larger choices of incremental lag $\tau$.
    The incremental time lags and best-fitting distributions are:
    CTH -- ($\Box$) $\tau=0.06\,$s, $f_{\mathrm{log}\mathcal{N}}(\beta)$; (\raisebox{-0.5mm}{\scalebox{1.5}{$\circ$}}) $\tau=0.30\,$s $f_{\Gamma}(\beta)$; (\rotatebox[origin=c]{45}{$\Box$}) $\tau=1.20\,$s $f_{\mathrm{log}\mathcal{N}}(\beta)$.
    LTU -- ($\Box$) $\tau=0.06\,$s, $f_{F}(\beta)$; (\raisebox{-0.5mm}{\scalebox{1.5}{$\circ$}}) $\tau=0.70\,$s $f_{F}(\beta)$; (\rotatebox[origin=c]{45}{$\Box$}) $\tau=1.20\,$s $f_{F}(\beta)$.
    TTY -- ($\Box$) $\tau=0.06\,$s, $f_{\mathrm{inv}\Gamma}(\beta)$; (\raisebox{-0.5mm}{\scalebox{1.5}{$\circ$}}) $\tau=0.16\,$s $f_{F}(\beta)$; (\rotatebox[origin=c]{45}{$\Box$}) $\tau=1.20\,$s $f_{F}(\beta)$.
    AU -- ($\Box$) $\tau=0.06\,$s, $f_{\mathrm{inv}\Gamma}(\beta)$; (\raisebox{-0.5mm}{\scalebox{1.5}{$\circ$}}) $\tau=0.38\,$s $f_{F}(\beta)$; (\rotatebox[origin=c]{45}{$\Box$}) $\tau=1.20\,$s $f_{\mathrm{inv}\Gamma}(\beta)$.
    See Tabs.~\ref{table} and \ref{table_2} for the exact Kolmogorov--Smirnov statistics at each location.}\label{fig:more_betas}
\end{figure*}

We have thus uncovered a time scale separation: at very short incremental lags $\tau < 1.2\,$s a superstatistical study is justified ($d \ll T$).
At incremental lags $\tau > 1.2\,$s this clear time scale separation ceases to exist.
In the following we will present our numerical results for the probability density $f(\beta)$ at very short lags $\tau = 0.06\,$s, 
at large lags $\tau = 1.20\,$s, and for four intermediate choices indicated in Fig.~\ref{fig:Ts} by circles, $\tau_{\mathrm{CTH}} = 0.30\,$s, $\tau_{\mathrm{LTU}} = 0.70\,$s, $\tau_{\mathrm{TTY}} = 0.16\,$s, and $\tau_{\mathrm{AU}} = 0.38\,$s.
These are the inflection points observed in Fig.~\ref{fig:Ts}.
In Fig.~\ref{fig:more_betas} we display the results.
LTH and KTH are not presented since they exhibit a varying kurtosis and strong correlation, and a somewhat atypical
behaviour.

As mentioned before, we cannot unambiguously distinguish between different theoretically possible $f(\beta)$ for the different time lags, as for some distributions the
Kolmogorov--Smirnov distance $D_n$ (or other metrics to determine the agreement of a fitting function with the data) are similar.
We note that $f(\beta)$ varies both with the time lag as well as with the location where the measurements are done.
Take for example the recordings at AU:
at $\tau=0.04\,$s and at $\tau=0.06\,$s $f(\beta)$ it resembles an inverse Gamma distribution; at $\tau=0.38\,$s $f_{F}(\beta)$ an $F$-distribution; at $\tau=1.20\,$s again an inverse Gamma distribution. Assuming this is a
a stable result and not just a statistical fluctuation,
this means there are transitions from one superstatistics
to another. 
This phenomenon of transitions between different types of superstatistics giving optimal fits to the data as a function of the time lag has been previously observed in Refs.~\cite{Xu2016,Jizba2018} for financial time series (share price differences) as a function of the time lag.

Next, we display for each of the four examined incremental time series correlation functions, both for the original incremental time series and as well as for the extracted volatility $\beta(t)$ (Fig.~\ref{fig:more_betas}, bottom panels).
From the physical point of view of superstatistics, the correlation of $\beta(t)$ must be longer than that of the original incremental time series, as we have previously discussed.
This guarantees that each incremental time series de-correlates faster than the changes in the superstatistical environment, ensuring an equilibrium is obtained locally before the system's physics changes.
This is verified for all cases.

Let us finally quantify the strength of the fluctuations of the volatilities $\beta$.
If there were no fluctuations of $\beta$, the distribution  would be sharply peaked, i.e., in \eqref{eq:fund_super_dist}, $f(\beta) = \delta(\beta - \langle\beta\rangle)$.
Following the original work \cite{Beck2003}, we quantify the width of the fluctuations of $\beta$ via the general entropic index $q$, defined for any superstatistics as
\begin{equation}\label{eq:q}
q=\frac{\langle \beta^2 \rangle}{\langle \beta \rangle^2},
\end{equation}
which evaluates the width of the variations of $\beta$.
As mentioned before, if we observe no variations in $\beta$, our data would not exhibit heavy tails and the entropic index would be just $q=1$.

In general one cannot analytically evaluate~\eqref{eq:fund_super_dist}.
Depending on the different distributions given in \eqref{eq:dist_lognormal}--\eqref{eq:dist_F}, one might or might not be able to solve the integral.
Nevertheless, for small fluctuations of $\beta$ the integral in \eqref{eq:fund_super_dist} can be expanded.
For small $\sigma^2 = \langle  \beta^2 \rangle - \langle  \beta \rangle^2$ we obtain
\begin{align}\label{eq:approx_f_N}
    p(\Delta f_\tau) &= p_{\mathcal{N}}(\Delta f_\tau|\langle\beta\rangle)\left[1 + \frac{1}{8}\sigma^2 \Delta f_\tau^4 +  \mathcal{O}(\sigma^3)\right]\\
    &= p_{\mathcal{N}}(\Delta f_\tau|\langle\beta\rangle)\left[1 + \frac{1}{8} (q-1)\langle\beta\rangle^2 \Delta f_\tau^4 +  \mathcal{O}(\sigma^3)\right],\nonumber
\end{align}
with $q$ as given in \eqref{eq:q}.
The expectations $\langle \beta^k \rangle$ in \eqref{eq:approx_f_N} are either directly formed with $f(\beta)$ (type-A superstatistics) or with a slightly deformed $\widetilde{f}(\beta) \sim \sqrt{\beta}f(\beta)$ (type-B superstatistics), see Ref.~\cite{Beck2003} for more details.
Again, if $q=1$, $p(\Delta f_\tau) = p_{\mathcal{N}}(\Delta f_\tau|\langle\beta\rangle)$ and we then expect no heavy tails in our increment distribution.

In Tab.~\ref{table_qs} the entropic indices $q$ for CTH, LTU, TTY, and AU, for three different incremental time lags, are shown.
The $q$-values do not change significantly with $\tau$, which is understandable given that the correlation times of the volatilities $\beta$ are far larger than the chosen  ranges of $\tau$.
What is noticeable is that each location has a very different entropic index $q$.
This tells us that the width of fluctuations of the volatilities at each location are very different in amplitude, a spatial heterogeneity for the power grid
on large spatial scales.
Whereas large $q$ are observed at CTH, indicating strong fluctuations in $\beta$, at AU these fluctuations are much smaller.
\begin{table}[t]
  \centering
  \begin{tabular}{r|c|c|c}
    \hline
    Location & $\tau=0.02$\,s & $\tau=0.04$\,s & $\tau=1.20$\,s \\ \hline
    CTH & 1.772 & 1.838 & 1.799 \\
    LTU & 1.572 & 1.618 & 1.568 \\
    TTY & 1.506 & 1.492 & 1.397 \\
    AU & 1.119 & 1.103 & 1.253 \\ \hline
  \end{tabular}
\caption{Entropic index $q$ for three different incremental lags $\tau=0.02$\,s, $\tau=0.04$\,s, and $\tau=1.20$\,s at CTH, LTU, TTY, and AU. We note here that the entropic index $q$ at each location does not change much whilst varying the incremental lag $\tau$, yet it is different for each location, indicating clearly the different nature of the fluctuations at each location.}\label{table_qs}
\end{table}

\section{Conclusion}
In this article we focused on examining the spatio-temporal complexity of the stochastic properties of six synchronous recordings of power-grid frequency in the Nordic Grid synchronous area, for some example time series measured in the year 2013.
{\em A priori} one would expect power-grid frequency to be essentially indistinguishable across a synchronous power grid due to the strong phase locking at play at each power generator in the power grid.
This nevertheless does not preclude stochastic fluctuations being present in the recordings---these are indeed ubiquitous and they exhibit complex behaviour on various temporal and spatial scales.

We have shown that the increment probability densities at six spatially distant locations show heavy tails which are quite different for each of these locations.
We introduced a superstatsitical model for this, considering the incremental statistics as arising from a superposition of Gaussian distributions via a superstatistical scaling function (a probability density of local variances).
We showed that there is time scale separation in the system (a necessary condition for superstatistics to work) for small time lags, and that there is a phase transition at incremental lags $\tau=1\sim2$ seconds where this description looses validity. This is in line with previous observations on the emergence of phase synchronisation in the same recordings~\cite{Forthcoming}. 
Moreover, we note that although we observe a fast decrease of the kurtosis as we increase each location's incremental time lag, the superstatistical properties remain very similar, i.e., the entropic indices $q$ remain constant.
This indicates that the decrease of the kurtosis for these recordings is not caused by a change in the internal properties of the system, it is more likely due to the strong phase locking between generator machines.

Although the data have a very high temporal resolution ($0.02$ seconds) they cover only $36$ hours of activities.
We nevertheless can quantify the strength of the fluctuations of the parameter $\beta$  of the underlying physical process by studying the entropic index $q$ of each recording,
which is a measure for the typical width of the fluctuations in $\beta$.
We observe that the entropic index $q$ remains largely constant in the range $<1.2$ seconds, yet it is different for each spatial location, indicating that the physics of the increments is highly inhomogeneous on a large spatial scale and dependent on the location of the recording in the power grid.
This indicates that the underlying physics of the increment statistics is different at each location, and these fluctuations are stronger in the southern part of the Nordic Grid---the more populated area of the Nordic grid.
This points to the need of having stronger local control mechanisms at points where more power is consumed.

From a superstatistical point of view, a particularly interesting result of our investigation is that the performed Kolmogorov--Smirnov tests point to different types of superstatistics being relevant at the six different locations, meaning there are different superstatistical distribution functions $f(\beta)$ giving the best fits to the data in the different regions. 
Of course, this spatial inhomogeneity should still be confirmed with bigger data sets and longer time series of other power grids in the future. 
Nevertheless, it is apparently a phenomenon that deserves further investigation, as it is a phenomenon more complex than for homogeneous hydrodynamic turbulence, where usually just one type of superstatistics (lognormal superstatistics) is sufficient.
Phase transitions from one superstatistics to another occur also as a function of the time lag for our data, this phenomenon has been previously observed for financial time series in Refs.~\cite{Xu2016,Jizba2018}.
Of course, some caution must be taken here, in the sense that a deeper statistical analysis for larger and longer data sets is necessary to confirm this result of different types of superstatistics
acting in a co-existing way.

To summarise, power-grid frequency, recorded synchronously across the Nordic Grid in our investigation, exhibits very complex spatio-temporal behaviour, if the small fluctuations around the mean are carefully taken into account by doing increment statistics.
On the one hand, each recording is very similar given the strong phase-locking within the power grid, on the other hand, the recorded power-grid frequency fluctuations follow different types superstatistics, depending on the spatial location, and depending on the time lag considered.

\paragraph*{Software} Numerical calculations and distribution fittings performed with \texttt{python'}s \texttt{SciPy}~\cite{SciPy} and \texttt{NumPy}~\cite{NumPy}. 
Figures generated with \texttt{Matplotlib}~\cite{Matplotlib}.

\paragraph*{Conflicts of Interest} The authors declare no conflict of interest.

\begin{acknowledgments}
We gratefully acknowledge support from the German Federal Ministry of Education and Research (grant no.~03EK3055B) and the Helmholtz Association (via the joint initiative ``Energy System 2050 -- A Contribution of the Research Field Energy'' and the grant ``Uncertainty Quantification -- From Data to Reliable Knowledge (UQ)'' with grant no.~ZT-I-0029).
This work was performed as part of the Helmholtz School for Data Science in Life, Earth and Energy (HDS-LEE). 
This project has received funding from the European Union's Horizon 2020 research and innovation programme under the Marie Skłodowska-Curie grant agreement No 840825.
\end{acknowledgments}

\bibstyle{apsrev4-2}
\bibliography{bib}

\end{document}